\documentclass[epj]{svjour}
\usepackage{graphics}
\usepackage{placeins}
\usepackage{graphicx}
\usepackage[square,numbers]{natbib}

\usepackage{graphicx}
\usepackage{placeins}
\usepackage{xcolor}
\usepackage{amsmath,amssymb}

\usepackage{verbatim}
\usepackage{multirow}
\usepackage{cuted}

\newcommand{\cl}{\mathcal{L}}
\newcommand{\ce}{\mathcal{E}}
\newcommand{\diff}{\mathrm{d}}

\begin{document}
\title{Epicyclic orbits in the field of Einstein-Dirac-Maxwell traversable wormholes applied to the quasiperiodic oscillations observed in microquasars and active galactic nuclei}
\author{Zden\v{e}k Stuchl\'{\i}k\inst{1}%
\thanks{\emph{Email address:} zdenek.stuchlik@physics.slu.cz}
\and Jaroslav Vrba\inst{1}
\thanks{\emph{Email address:} jaroslav.vrba@physics.slu.cz}%
}                     
%
%
\institute{$^\textrm{1}$Research Centre for Theoretical Physics and Astrophysics, Institute of Physics, Silesian University in Opava, Bezru\v{c}ovo n\'am.~13, 746\,01, Opava, CZ }
\date{Received: date / Revised version: date}
%
\abstract{
We study the epicyclic oscillatory motion around circular orbits of the traversable asymptotically flat and reflection-symmetric wormholes obtained in the Einstein-Dirac-Maxwell theory without applying exotic matter in their construction. We determine frequencies of the orbital and epicyclic motion in the Keplerian disks having inner edge at the marginally stable circular geodesic of the spacetime. The obtained frequencies are applied in the so called geodesic models of high-frequency quasiperiodic oscillations (HF QPOs) observed in microquasars and active galactic nuclei containing a supermassive central object. We show that even the simplest epicyclic resonance variant of the geodesic models can explain the HF QPOs observed in many active galactic nuclei for realistic choices of the wormhole parameters, but there are some of the sources where only wormholes with unrealistically large values of the parameters can be sufficient for the explanation. On the other hand, in the case of microquasars, the observed HF QPOs strongly restrict the acceptable values of the wormhole parameters. 
%
} 
\maketitle
\section{Introduction}
Wormholes are interesting and popular alternatives to the black holes representing a stabilized tunnel (throat) connecting different universes, or distant parts of the Universe. The standard traversable wormholes were introduced by Morris and Thorne \cite{Mor-Tho-Yur:1988:PhysRevLet:}, the simple reflection-symmetric traversable wormholes were studied by Visser \cite{Vis:1989:NuclPhysB:,Poi-Vis:1995:PHYSR4:} -- as two Swarzschild spacetimes connected by a shell constituted by extraordinary matter violating the weak energy condition because of negative energy density that guarantees validity of Einstein gravitational equations for wormholes stable against gravitational collapse. Later it was demonstrated that such extraordinary stress-energy tensor can be avoided in wormhole spacetimes constructed in high-dimensional general relativity \cite{Svi-Tah:2018:EPJC:}, or alternative versions of gravity theories \cite{Har-Lob-Mak:2013:PhRvD:}, or for fermions giving negative Casimir energy \cite{Mal-Mil-Pop:2018:arxiv:}. 

Of special character is recently popular meta-geometry introduced by Simpson and Visser \cite{Sim-Vis:2019:JCAP:}, giving a continuous transition (transcendence) between regular black hole states and wormhole states; note that a wormhole is hidden under the event horizon of the regular black hole. The rotational Kerr-like generalization of the meta-geometry was recently given in \cite{Maz-Fra-Lib:2021:JCAP:}. Interesting phenomena were found in studies of the Simpson-Visser meta-geometry for various astrophysical phenomena \cite{Zho-Xie:2020:EPJC:,Stu-Vrb:2021:Universe:}. 

The astrophysical phenomena were studied for the traversable wormholes in many papers, related mainly to optical effects as the shadow \cite{Abd-Jur-Ahm:2016:ASS:,Bou-Che-Che:2021:arxiv:}, or weak and strong lensing. Note that the shadow of wormhole means the shape of the photon sphere rather than a black spot cast by the black hole; however, if in an appropriate limit of the wormhole parameters they approach a black hole state, near the transition point the wormholes mimic black hole behavior \cite{Dam-Sol:2007:PRD:,Car-Fra-Pan:2016:PhysRevLet:,Chur-Stu:2020:CLAQG:,Bro-Kon-Pap:2021:PHYSR4:}. 

High astrophysical interest is related to the models of oscillatory phenomena observed near compact objects in microquasars and active galactic nuclei \cite{Rem-McCli:2005:ARAA:}. Of special interest are so called geodesic model of twin high-frequency quasiperiodic oscillations observed in microquasars and active galactic nuclei \cite{Abr-Klu:2001:AAP:,Tor-etal:2005:AA:,Stu-Kot-Tor:2013:ASTRA:,Stu-etal:2020:UNI:} that seem to be very successful in the case of microquasars \cite{Tor-etal:2011:ASTRA:,Stu-Kol:2016:ASTRA:} but seem to be irrelevant for data observed in active galactic nuclei, if we use the standard assumption of central supermassive black holes \cite{Smi-Tan-Wag:2021:ApJ:}. Therefore, it is of high interest, if this fault of the standard HF QPOs geodesic models could be cured by assumption of a central supermassive wormhole instead of a black hole \cite{Stu-Vrb:2021:Universe:,Del-Kun-Ned:2021:arxiv:,Del-etal:2021:PRD:}. 

Quite recently, a new physically very interesting family of traversable wormholes was found in the framework of the Einstein-Dirac-Maxwell theory. Contrary to the cases where the wormholes with no exotic matter are constructed in modifications of general relativity, in the later case, the asymptotically flat symmetric traversable wormholes are constructed in standard general relativity enriched by Maxwell and two Dirac fields with standard coupling \cite{Bla-Kno-Rad:2021:PhRvL:,Bla-Sal-Kno-Rad:2021:arxiv:} -- a modification improving some issues related to the original model was proposed in \cite{Kon-Zhi:2021:arxiv:}. Quasinormal modes, echoes and shadows of these new wormhole solutions were studied in \cite{Chu-Kon-Stu:2021:arxiv:}. 

In the present paper we consider the possibility to explain the HF QPOs observed in active galactic nuclei by proper choice of the parameters of the Einstein-Dirac-Maxwell wormhole geometry. We also test limits on these wormhole parameters determined by the fitting geodesic models to the data observed in microquasars. To reach these goals, we determine the frequencies of the orbital and epicyclic motion of test particles, relevant for Keplerian disks, and apply them in the epicyclic resonance variant \cite{Tor-etal:2005:AA:} of the geodesic models of HF QPOs.

\section{Einstein-Dirac-Maxwell wormhole geometry}
The Einstein-Dirac-Maxwell (EDM) wormhole spacetime along with the related electromagnetic structures was derived in \cite{Bla-Kno-Rad:2021:PhRvL:}. The analytical solution can be found for massless spinor fields and complete form of this solution can be found in \cite{Chu-Kon-Stu:2021:arxiv:}. As we consider in our paper only interaction of this solution with electrically neutral matter, we give here only the form of the spacetime line element that can be expressed in the form
\begin{equation}
	\diff s^2=-\left(1-\frac{2M}{r}\right)^2\diff t^2 + \frac{\diff r^2}{\left(1-\frac{r_0}{r} \right)\left(1-\frac{Q_e^2}{r_0\,r} \right)} + r^2\left(\diff\theta^2+\sin^2\theta\,\diff\phi^2\right)
	\label{e:met}
\end{equation}
where the mass parameter can be expressed as 
\begin{equation}
    M=\frac{Q_e^2\,r_0}{Q_e^2+r_0^2}.
\end{equation}
This spacetime describes a traversable wormhole having radius of the throat given by the parameter $r_0$ and electric charge satisfying the condition $|Q_e| < r_0$. 
The parameters of the metric including $M$ are not chosen freely, they have to obey the following condition  \cite{Bla-Kno-Rad:2021:PhRvL:}:
\begin{equation}
    M<Q_e<r_0.
\end{equation}

The wormhole is stabilized by the spinor contribution to the total stress-energy tensor that is regular at all points of the spacetime. For $Q_e \to r_0$ the wormhole spacetime approaches the extremal Reissner-Nordstrom black hole spacetime -- for more details see e.g. \cite{Chu-Kon-Stu:2021:arxiv:}. 

For simplification we introduce dimensionless wormhole parameters and radial coordinate by the transformations 
\begin{equation}
    Q_e \rightarrow \frac{Q_e}{M}, \quad r_0 \rightarrow \frac{r_0}{M}  \quad \textrm{and} \quad r \rightarrow \frac{r}{M}.
\end{equation}

\section{Motion of test particles and the circular orbits}
The motion of test particles is governed by the spacetime geometry, namely by its geodesics that are timelike for test particles with non-zero rest mass, and null for photons \cite{Mis-Tho-Whe:1973:Gravitation:}. The spacetime symmetries imply existence of two constants of motion
\begin{equation}
    E = -p_t \qquad L = p_{\phi},
    \label{e:conserv}
\end{equation}
energy $E$ and axial angular momentum $L$. Due to the spherical symmetry of the geometry, the motion is allowed in central planes only. For simplicity we can chose for a selected particle the equatorial plane $\theta=\textrm{const}=\pi/2$).

Hereafter, we focus on circular geodesics, which can be studied by examining the effective potential, which in the case of EDM wormholes takes the form:
\begin{equation}
    V_\mathrm{eff}=\frac{(r-2)^2 \left(L^2-\epsilon\, r^2\right)}{r^4},
    \label{e:efpot1}
\end{equation}
where $L$ is angular momentum defined (\ref{e:conserv} and $\epsilon$ distinguishes massive ($\epsilon=-m$) and massless ($\epsilon=0$) test particles.

The parameters of the EDM wormhole does not affect the behaviour of the effective potential in the usual sense. They only change the mass of the central object. For example, it does not change the position of the circular orbits, however, $r_0$ changes the radius of the horizon, i.e. the specific orbit can be hidden below the horizon if $r_0$ is greater than the radius of this orbit. This is a difference to Schwarzschild black hole spacetime, where the horizon is fixed at $r=2M$.

\subsection{Photons}
\begin{figure}[ht] 
    \centering
	\includegraphics[width=0.5\linewidth]{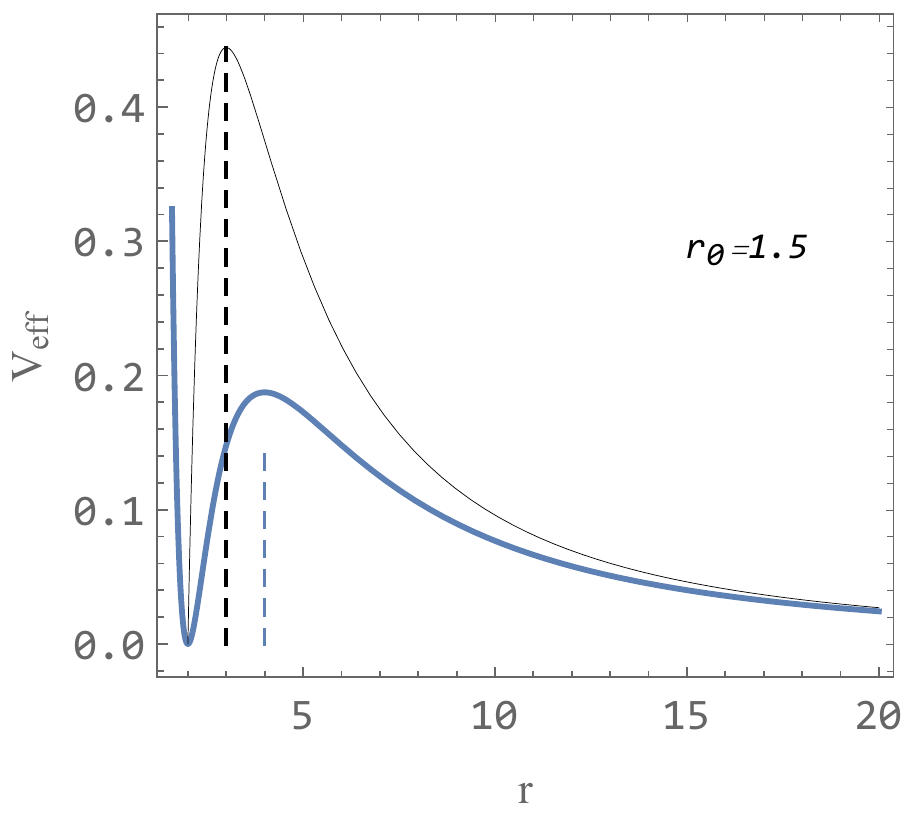}
	\caption{Effective potential of photons in the field of the EDM wormhole (blue) compared to Schwarzschild's effective potential (black).}
	\label{f:f1}
\end{figure}

The motion of the photons is independent of its energy and photons move along null-geodesics. In the effective potential (\ref{e:efpot1}) we put $\epsilon=0$ and the comparison to the Schwarzschild effective potential is shown in Fig. \ref{f:f1}. Notice the shape of the effective potential would not change if we change $r_0$ or $Q_e$. The change of the $r_0$ would cause a change of the beginning of the curve of the effective potential (the beginning of the outer spacetime).
The circular photon orbits $r_\mathrm{cpo}$ are situated at the extrema of the effective potential which satisfy the condition $\diff V_\mathrm{eff}/\diff r=0$. It gives
\begin{equation}
    \diff V_\mathrm{eff}/\diff r=-\frac{2 L^2 (r-4) (r-2)}{r^5}=0.
\end{equation}
One immediately sees that for $r_0<2$ there is an inner stable photon circular orbit located at 
\begin{equation}
    r_\mathrm{cpo(s)}= 2 
\end{equation}
and an outer unstable photon circular orbit located at 
\begin{equation}
    r_\mathrm{cpo(u)}= 4.
\end{equation}
This behavior is in clear contrast with the Schwarzschild case where only the unstable circular photon orbit is located at $r_\mathrm{cpo}=3$. For $r_0>2$ only the unstable photon circular orbit is relevant -- notice significant shift of this orbit with respect to the Schwarzschild black hole spacetimes. 

Note that the shadow of the wormhole is governed by the impact parameter ($l=L/E$) of the unstable photon circular orbit that takes the value 
\begin{equation}
    l_\mathrm{cpo(u)}= 8.
\end{equation}
The Wormhole shadow is thus independent of the parameter $r_0$ being larger than in the case of the Schwarzschild black holes -- this is a natural result as the limiting value of the parameter $r_0$ corresponds to the extreme Reissner-Nordstrom black hole, not the Schwarzschild black hole. 

The stable photon circular orbits located at $r_\mathrm{cpo}=2$ imply a region of trapped null geodesics (photons) in the region extending up to the radius $r_\mathrm{cpo(u)}= 4$. This situation corresponds to the existence of regions of trapped null geodesics in the field of Kerr (Kerr-Newman) naked singularity \cite{Stu-Sche:2010:CLAQG:,Stu-Sche:2013:CLAQG:,Bla-Stu:2016:PhRvD:,Stu-Bla-Sche:2017:PhRvD:} or discovered in a variety of compact objects, see e.g. \cite{Stu-Hle-Nov:2016:PRD:,Nov-Stu-Hla:2017:PRD:}. An instability against gravitational perturbations was demonstrated in the trapping region of trapping polytropes \cite{Stu-Sch-Tos:2017:JCAP}.

\subsection{Massive test particles}
In the case of the massive test particles ($\epsilon=-m$), it is convenient to introduce new constants of motion related to the rest mass of the particle, namely the specific angular momentum and specific energy 
\begin{equation}
    \cl= \frac{L}{m} \quad \textrm{and}\quad \ce = \frac{E}{m} . 
\end{equation}
The effective potential of the motion of the massive test particles then takes the form having the specific angular momentum as a parameter 
\begin{equation}
    V_\mathrm{eff}=\frac{(r-2)^2 \left(\cl^2+ r^2\right)}{r^4}.
    \label{e:efpot2}
\end{equation}

\begin{figure}[ht] 
    \centering
	\includegraphics[width=0.5\linewidth]{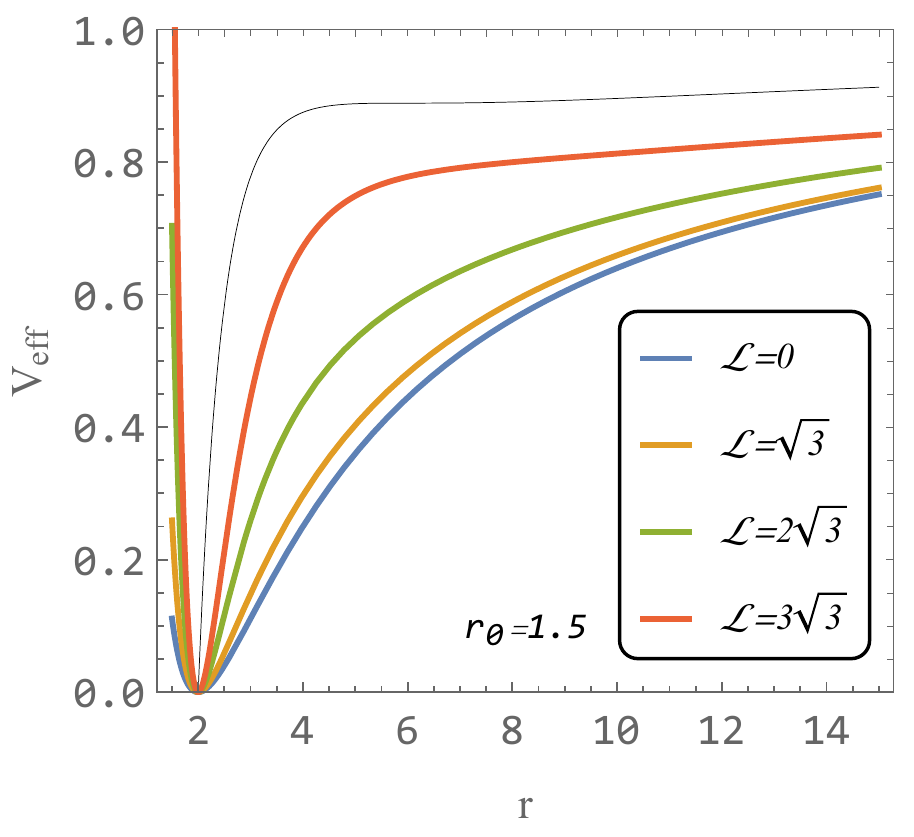}
	\caption{Effective potential of massive test particles in the field of the EDM wormhole compared to Schwarzschild's effective potential (black).}
	\label{f:f2}
\end{figure}

Comparison with the Schwarzschild effective potential is given at Fig. \ref{f:f2}.

\begin{figure}[ht] 
    \centering
	\includegraphics[width=0.5\linewidth]{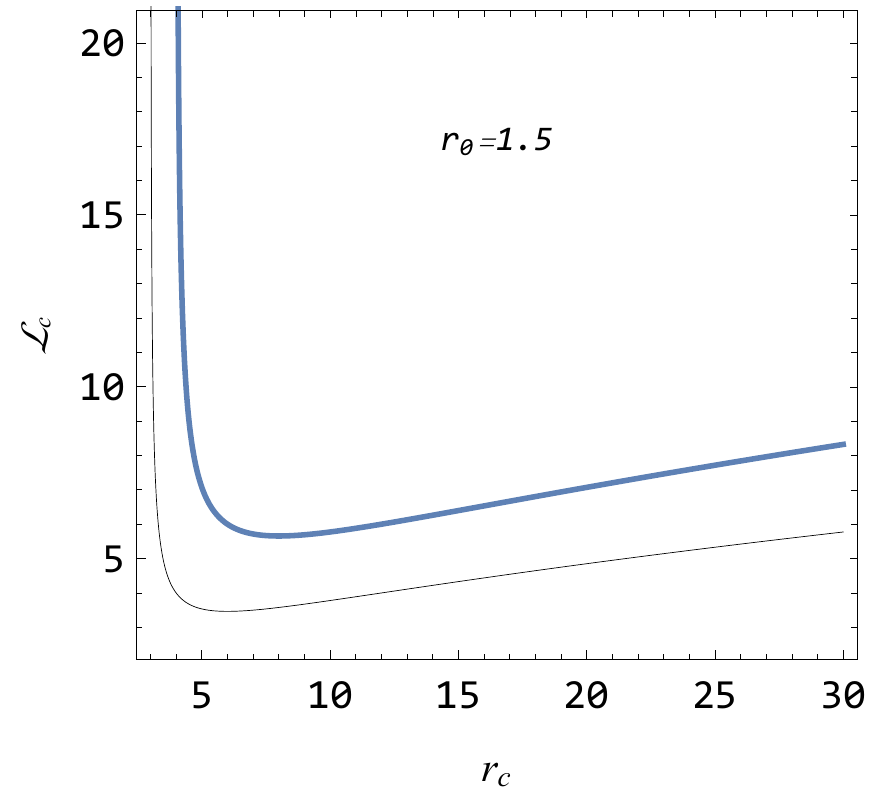}
	\caption{The angular momentum at the circular orbit of in the field of the EDM wormhole (blue) compared to Schwarzschild's angular momentum at the circular orbit (black).}
	\label{f:f3}
\end{figure}

The circular orbits are governed by the local extrema of the effective potential, i.e. by the condition  
\begin{equation}
    \diff V_\mathrm{eff}/\diff r=\frac{2 (r-2) \left(2 r^2-\cl^2 (r-4)\right)}{r^5}=0.
    \label{e:dveff}
\end{equation}
Eq. (\ref{e:dveff}) can be analytically solved for $\cl$. The radial profile of the specific angular momentum of the circular geodesics $\cl_\mathrm{c}(r)$ is then given by the relation 
\begin{equation}
    \cl_\mathrm{c}=r\sqrt{\frac{2}{r-4}}.
    \label{e:lc}
\end{equation}
There is another point of vanishing of the effective potential at $r=2$, corresponding to the stable circular photon orbit, as demonstrated in discussion of the photon motion. 
One immediately sees that there are no other circular orbits at $r\leq4$ and at $r=4$ where the radial profile diverges the unstable photon circular orbit is located. The comparison with the radial profile of the angular momentum in the Schwarzschild spacetime is demonstrated in Fig. \ref{f:f3}. 

The circular orbits are stable only at radii larger than the radius of the innermost stable circular orbit (ISCO). To find the ISCO, we have to solve simultaneously equations $\diff V_\mathrm{eff}/\diff r=0$ and $\diff^2 V_\mathrm{eff}/\diff r^2=0$, where
\begin{equation}
    \diff^2 V_\mathrm{eff}/\diff r^2=\frac{\cl^2 \left(6 r^2-48 r+80\right)-8 (r-3) r^2}{r^6}. 
\end{equation}
Using the radial profile of the specific angular momentum, the analytic solution reads $r_\mathrm{ISCO}=8$. Clearly, this is significantly larger than in the Schwarzschild spacetime where there is $r_\mathrm{ISCO}=6$.

\section{Frequencies of epicyclic motion}
A test particle orbiting at a stable circular orbit, $r_c$, in the equatorial plane may be slightly displaced, starting oscillatory epicyclic motion around the $r_c$ and the latitude $\theta=\pi/2$. Displacement for a small perturbations in the radial direction may by written as $r=r_c+\delta r$, and in the latitudinal direction as $\theta=\pi/2+\delta\theta$. In the linear perturbative regime, equations giving the radial and latitudinal epicyclic motion around are equivalent to the harmonic oscillator equations \cite{Tor-Stu:2005:ASTRA:,Stu-Kot-Tor:2013:ASTRA:}
\begin{eqnarray}
\delta\ddot{ r}+\bar{\omega}_r^2 \delta r = 0, \ \ \ 
\delta\ddot{ \theta}+\bar{\omega}_\theta^2 \delta\theta = 0.
\end{eqnarray}
where $\bar{\omega_{r}}$ ($\bar{\omega_{\theta}}$) represents the angular velocity of the radial (latitudinal, or vertical) epicyclic oscillations as measured at the radius of the circular orbit by the local observer.

The orbital angular frequency of the circular motion is determined directly by the equation of motion and take the form  
\begin{equation}
\bar{\omega}_\phi = \frac{\cl}{g_{\theta\theta}}.
\end{equation}

The angular frequencies of the radial and latitudinal epicyclic motion can be determined in the framework of the Hamiltonian formalism \cite{Kol-Stu-Tur:2015:CLAQG:,Tur-Stu-Kol:2016:PHYSR4:,Stu-Kol:2016:EPJC:,Kol-Tur-Stu:2017:EPJC:}. The Hamiltonian is defined as 
\begin{eqnarray}\label{e:ham}
H=\frac{1}{2}g^{\alpha\beta}p_\alpha p_\beta+\frac{m^2}{2}.
\end{eqnarray}
and can be separated into the dynamic and potential parts 
\begin{eqnarray}
H=H_{\mathrm{dyn}}+H_{\mathrm{pot}},
\end{eqnarray}
where
\begin{eqnarray}
H_{\mathrm{dyn}}&=& \frac{1}{2}\Big(g^{rr}p_r^2 +g^{\theta\theta}p_\theta^2\Big),\\
H_{\mathrm{pot}}&=& \frac{1}{2}\Big(g^{tt}\mathcal{E}^2+g^{\phi\phi}\cl^2+1\Big).\label{e:hpot}
\end{eqnarray}
In the standard way, the potential part of the Hamiltonian governs the locally defined radial and latitudinal epicyclic angular frequencies $\bar{\omega}_r$ and $\bar{\omega}_\theta$  
\begin{eqnarray}
\label{e:bom} 
\bar{\omega}_r^2 &=& \frac{1}{g_{rr}}\frac{\partial^2 H_{\mathrm{pot}}}{\partial r^2},\\ \nonumber
\bar{\omega}_\theta^2 &=& \frac{1}{g_{\theta\theta}}\frac{\partial^2 H_{\mathrm{pot}}}{\partial \theta^2}.
\end{eqnarray}
Using the relations (\ref{e:met}), (\ref{e:lc}), and (\ref{e:hpot}) in (\ref{e:bom}) and putting for simplicity $M=1$, we arrive at the formula for the epicyclic frequencies and the azimuthal frequency (as related to the local observers) in the form 
\begin{eqnarray}
\label{e:bomegas1}
\bar{\omega}_\theta &=& \bar{\omega}_\phi = \sqrt{\frac{2}{\left(r-4\right) r^2}},\\ \nonumber
\bar{\omega}_r &=&  \sqrt{\frac{2(r-8) (r-r_0) \left(r\,r_0 -Q_e^2\right)}{r^4 \,r_0 \left(r^2-6 r+8\right)}}.
\end{eqnarray}
As usual in the spherically symmetric spacetimes, the angular frequency of the latitudinal epicyclic oscillations coincide with the angular (azimuthal) frequency of the orbital motion. It should also be emphasized that these two frequencies are independent of the wormhole parameters $r_0$ and $Q_e$, i.e., they are identical for any combination of these two parameters. However, the radial epicyclic frequency depends on both these parameters. 
\begin{figure}[ht] 
	\includegraphics[width=\linewidth]{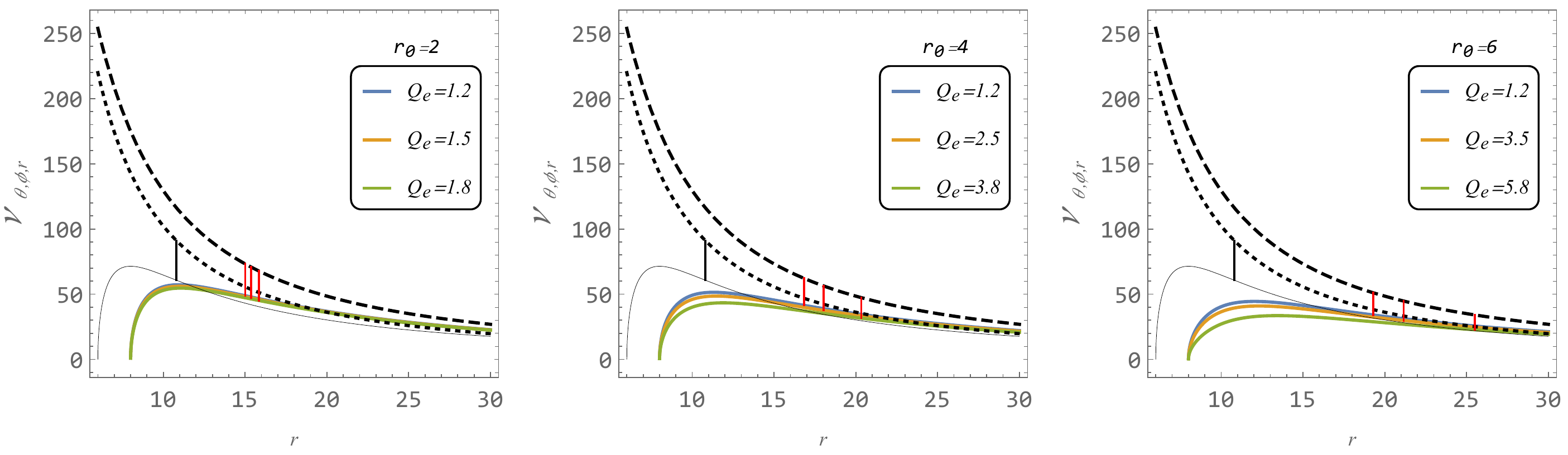}
	\caption{Epicyclic frequencies of the orbital resonance model in the field of the EDM wormhole (radial colored, latitudinal/angular black dashed) compared to Schwarzschild's one (radial black thin, latitudinal/angular black dotted). The red vertical lines indicates frequency ratio 3:2 in EDM wormhole spacetime and black vertical line in Schwarzschild spacetime.}
	\label{f:f4}
\end{figure}

However, observationally relevant are the angular frequencies measured by static observers at infinity, or very large distances, who represent real observers. For this reason, we transform the locally measured angular frequencies using the redshift factor corresponding to the orbital motion along the stable circular geodesic at $r_c$. We thus use the transform formula 
\begin{eqnarray}\label{e:transom}
\omega=\frac{\bar{\omega}}{-g^{tt}\mathcal{E}}, 
\end{eqnarray}
where we apply the metric coefficient $-g^{tt}$ and the covariant energy $\mathcal{E}$ taken at the radius of the stable orbit. Then the frequencies related to the distant static observers, and expressed in the standard units, are given by the relation 
\begin{equation}
    \nu_{i}=\frac{1}{2\pi}\frac{c^3}{G\,M}\frac{\bar{\omega}_{i}}{-g^{tt}\mathcal{E}},
    \label{e:transfreq}    
\end{equation}
where $i = {r, \theta, \phi}$. These expressions could be directly used in the fitting to observational data. The radial profiles of these frequencies are presented in Fig. \ref{f:f4}.

The radial epicyclic frequency is affected mostly by the change in the parameter $Q_e$. As seen from Fig. \ref{f:f4} the rising of parameter $Q_e$ lowers the upper frequency used for fitting of observation data from microquasars and supermassive active nuclei under the value in Schwarzschild case.

The HF QPOs in black hole are usually detected with the twin peaks which have a frequency ratio - upper:lower - close to 3:2. Therefore, in the subsequent modeling of epicyclic frequencies, we will emphasize this ratio and study how it is affected by the wormhole parameters $r_0$ and $Q_e$. 

The geodesic models can be represented by a large variety of variants where the observed upper $\nu_u$ and lower $\nu_l$ frequencies are related to some combinations of the epicyclic frequencies and the orbital frequency \cite{Stu-Kot-Tor:2013:ASTRA:}. Here we concentrate our discussion on the simplest and very important variants that could work in some sense in opposite way, as we shall see in the following discussion. 

\subsection{Epicyclic resonance model} 
This variant of the model was successfully introduced in modeling the observed frequencies in the known microquasars \cite{Tor-etal:2005:AA:}. In this variant the correspondence between the theoretical frequencies and the observed frequencies is assumed in the form 
\begin{eqnarray}
\label{e:bomepi}
\nu_{u} &=& \nu_\theta \\ \nonumber
\nu_{l} &=& \nu_r.
\end{eqnarray}
The case of the parametric resonance could then well explain the observed $3:2$ frequency ratio. Position of this frequency relation is then given by direct ratio of the epicyclic latitudinal and radial frequencies, as seen on Fig. \ref{f:f4}. Note that the epicyclic resonance model could work well both for a radiating hot spot, or radiating slender torus where the radial and vertical oscillations have the same frequencies as test particles under the epicyclic motion \cite{Rez-Yos-Zan:2003:MNRAS:}. 

\subsection{Relativistic precession model}
The relativistic precession model was introduced in \cite{Ste-Vie:1999:ApJ:} and is directly related to the precession motion of a particle in the equatorial plane. The upper frequency is thus directly given by the Keplerian frequency of the orbital motion. The correspondence of the theoretical frequencies to the observed frequencies is now given by the relations 
\begin{eqnarray}
\label{e:bomrel}
\nu_{u} &=& \nu_\phi = \nu_K \\ \nonumber
\nu_{l} &=& \nu_K - \nu_r.
\end{eqnarray}

\begin{figure}[ht] 
	\includegraphics[width=\linewidth]{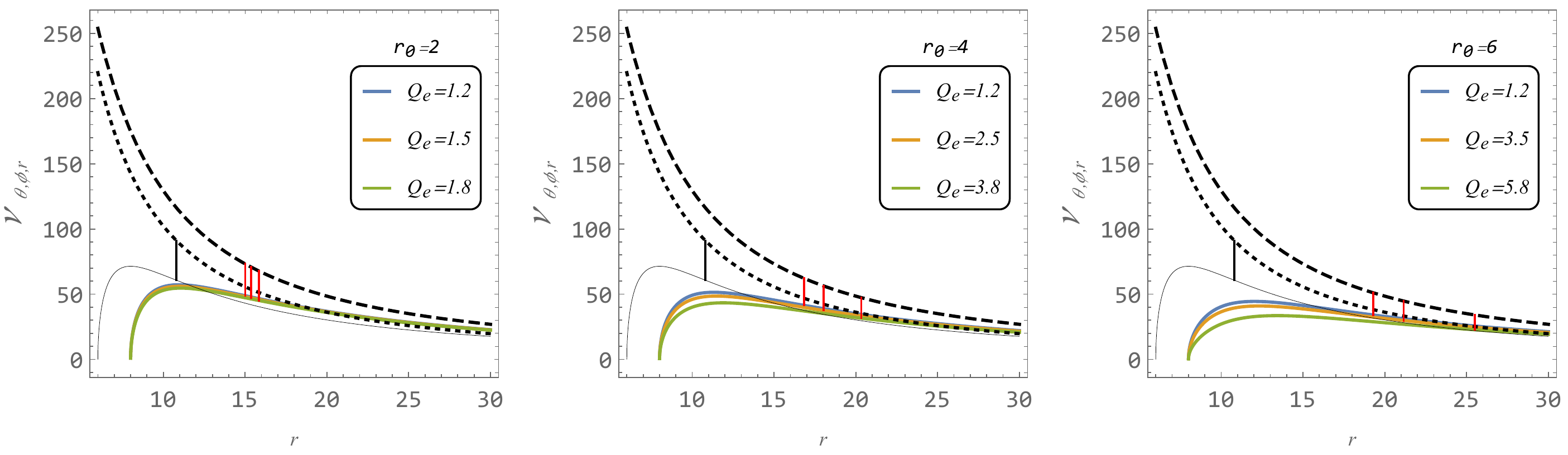}
	\caption{Epicyclic frequencies of relativistic precession model in the field of the EDM wormhole (radial colored, latitudinal/angular black dashed) compared to Schwarzschild's one (radial black thin, latitudinal/angular black dotted). The red vertical lines indicates frequency ratio 3:2 in EDM wormhole spacetime and black vertical line in Schwarzschild spacetime.}
	\label{f:f5}
\end{figure}

The Keplerian frequency $\nu_\mathrm{K}$ is independent of the wormhole parameters $r_0$ and $Q_e$, as the latitudinal frequency, while however, the radial epicyclic frequency is dependent on both wormhole parameters. Radial profiles of the resulting frequencies of the relativistic model are presented in Fig. \ref{f:f5}. 
In this model, the upper frequency of the 3:2 frequency ratio is higher than the Schwarzschild one for any of the applied parameters. The relativistic precession model thus works in opposition to the epicyclic resonance model, as seen in Fig. \ref{f:f5}. 

\section{Epicyclic resonance and relativistic precession variants of the geodesic model applied for HF QPOs observed in microquasars and around active galaxy nuclei}
The HF QPOs are of high importance in astrophysics, as they give well established restrictions on the parameters of the black holes (or alternative compact objects) in microquasars representing binary systems containing a stellar mass black hole, or in active galactic nuclei having supermassive black holes. Frequencies of the HF QPOs are measured in hundreds of Hz in the case of microquasars, but they are observed by six to ten orders smaller around the supermassive black holes. Due to the inverse-mass scaling of the observed frequencies, typical for the relations governing the epicyclic frequencies of the orbital motion \cite{Rem-McCli:2005:ARAA:}, the models of HF QPOs based on the orbital motion and related epicyclic oscillations, the geodesic models of HF QPOs determined by combinations of these frequencies, are very promising. Namely because the HF QPOs are often observed in the rational ratio \cite{McCli-Rem:2006:BHbinaries:}, especially in the ratio 3:2 \cite{Tor-etal:2011:ASTRA:}, indicating presence of resonant phenomena \cite{Klu-Abr:2001:ACTAASTR:}. Properties of the geodesic models are summarized in \cite{Stu-Kot-Tor:2013:ASTRA:} Inclusion of an electromagnetic interaction between a charged oscillating object and a magnetized black hole can be found in  \cite{Kol-Tur-Stu:2017:EPJC:}. In the geodesic models, the upper $\nu_\mathrm{u}$ and lower $\nu_\mathrm{l}$ observed frequencies are assumed to be a combination of the orbital and epicyclic frequencies (these frequencies are relevant for oscillations of slender tori \cite{Rez-Yos-Zan:2003:MNRAS:} too). 

The fitting to the observational data given for the microquasars and the active galactic nuclei \cite{Smi-Tan-Wag:2021:ApJ:} is realized for both the epicyclic resonance and relativistic precession model that behave from the point of view of the fitting procedure in opposite directions. The fitting curves are compared to the results of the fitting procedures realized for Kerr black holes -- we give the limiting fitting curves corresponding to the Schwarzschild case with dimensionless spin $a=0$, and the extreme Kerr black holes with $a=1$. The epicyclic resonance variant predicts fitting lines under the Schwarzschild line, while the relativistic precession variant predicts these lines above the Schwarzschild limit. 

We construct the fitting curves for the chosen variants of the geodesic model by the method using the inverse mass scaling valid for all the orbital and epicyclic frequencies and their combinations, as introduced in \cite{Stu-Kot-Tor:2013:ASTRA:}. Observational data corresponding to the HF QPOs around compact objects in active galactic nuclei are taken from \cite{Smi-Tan-Wag:2021:ApJ:} and summarized in Table \ref{t:tab1}.

\subsection{Microquasars fitting by the epicyclic resonance model}
We give the region of fitting by the Kerr black hole spacetimes using the epicyclic resonance variant of the geodesic model, limited by the cases $a=0$ ad $a=1$ where we introduce the limits on the mass parameter of the black holes (central objects) as established for the microquasars by methods different from the frequency measurements; usually in mass estimates the weak field limit approximation is used. The fitting lines determined by the epicyclic resonance variant constructed for the EDM wormholes are presented for reasonably chosen values of the two free parameters. Namely, the parameter $r_0$ is limited by value $6$, the values of the electric parameter are limited by the value of $r_0$. We can see that the all the wormhole fitting lines are generally located under the Schwarzschild fitting lines, excluding thus possibility of positive fitting for microquasars.   
\begin{figure}[ht] 
	\includegraphics[width=\linewidth]{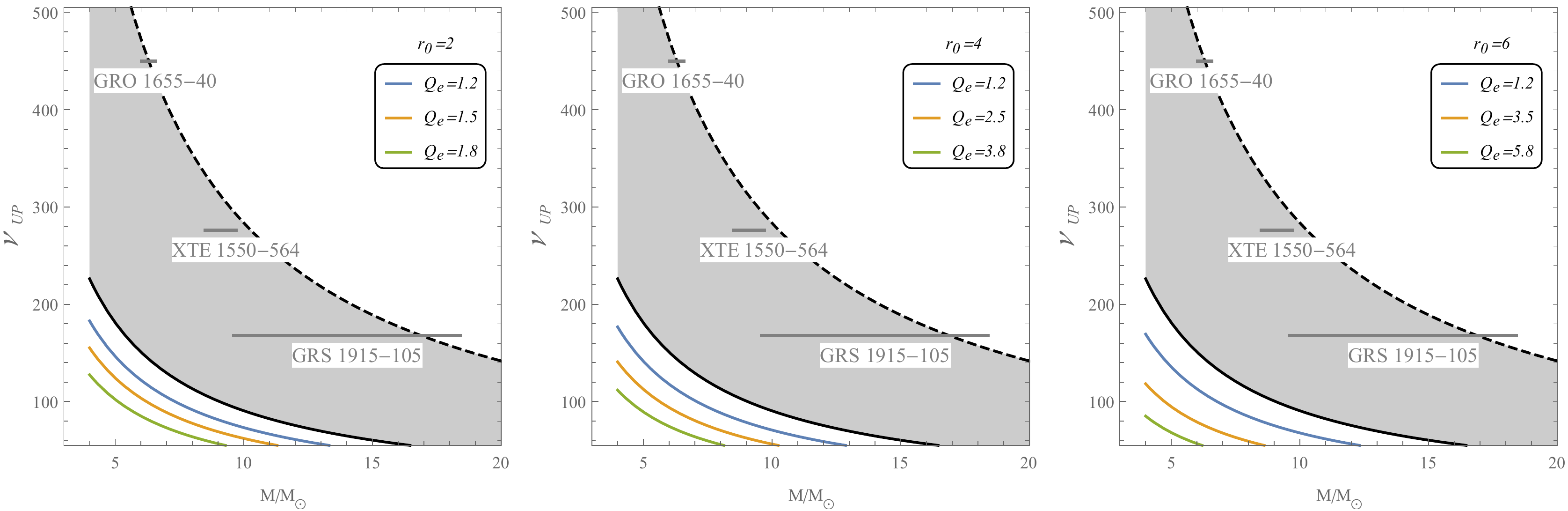}
	\caption{Fit of upper frequencies of 3:2 ratio to mass estimate of microquasars for various parameters $r_0$ and $Q_e$ by using the epicyclic orbital resonance model.}
	\label{f:f6}
\end{figure}
\begin{figure}[ht] 
	\includegraphics[width=\linewidth]{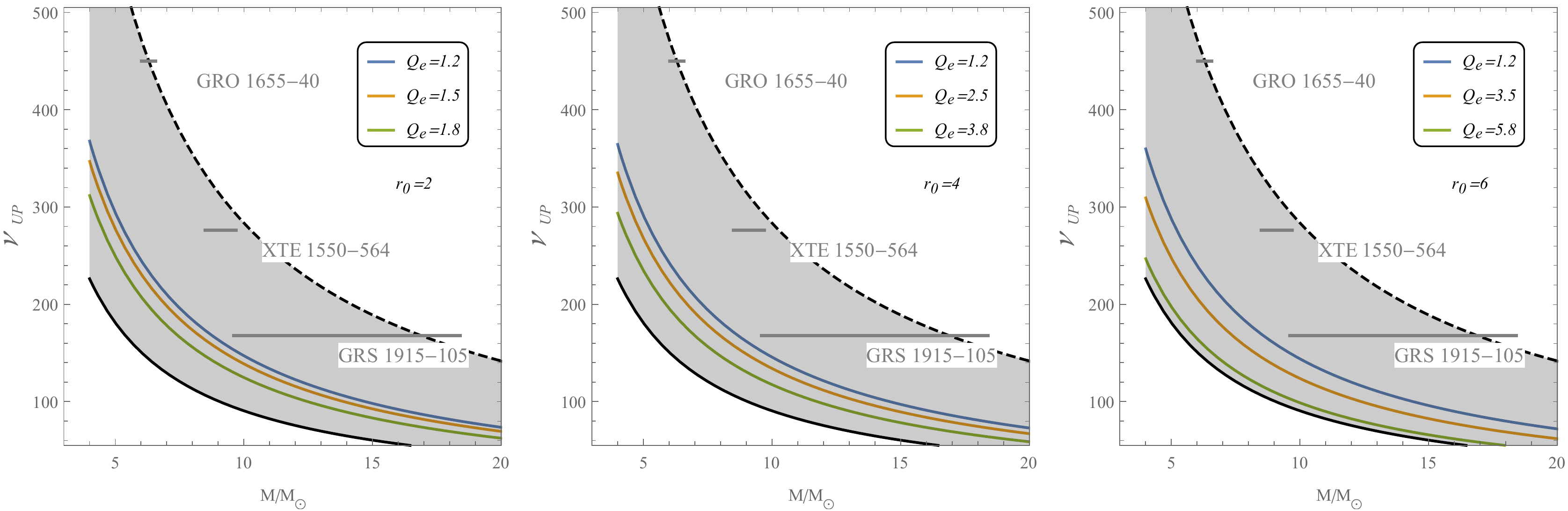}
	\caption{Fit of upper frequencies of 3:2 ratio to mass estimate of microquasars for various parameters $r_0$ and $Q_e$ by using the epicyclic relativistic precession model.}
	\label{f:f7}
\end{figure}

\subsection{Active galactic nuclei fitting by the epicyclic resonance model}
In the case of the fitting of data from the active galactic nuclei we use the same procedure and the same values of dimensionless wormhole parameters as in the case of the microquasars. In this case the character of the fitting lines related to wormholes enables positive fitting in the case of some (but not all) active galactic nuclei, as the data are located under the region of fitting by the Kerr black holes. 
\begin{figure}[ht] 
	\includegraphics[width=\linewidth]{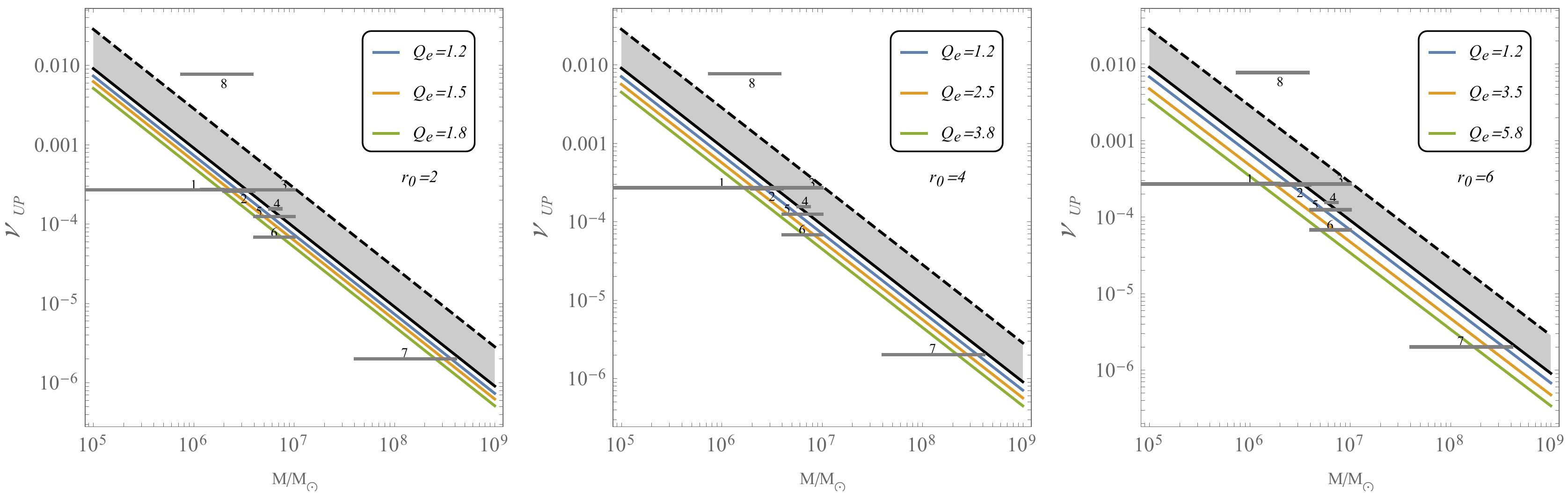}
	\caption{Fit of upper frequencies of 3:2 ratio to mass estimate of assumed supermassive black holes in active galactic nuclei for various values of the parameters $r_0$ and $Q_e$ by using the epicyclic orbital resonance model. The sources with known estimate of spin listed in Table \ref{t:tab1} (cf.  \cite{Smi-Tan-Wag:2021:ApJ:}).}
	\label{f:f8}
\end{figure}
\subsection{Microquasars fitting by the relativistic precession model}
We give the region of fitting by the Kerr black hole spacetimes due to the relativistic precession variant of the geodesic model, limited by the cases $a=0$ ad $a=1$ where we introduce the limits on the mass parameter of the black holes (central objects) as established for the microquasars by methods different from the frequency measurements. The fitting lines determined by the relativistic precession variant constructed for the EDM wormholes are presented for reasonably chosen values of the two free parameters chosen in the same way as in the case of the epicyclic resonance variant. We can see that in this case, contrary to the case of the epicyclic resonance variant, all the wormhole fitting lines are generally located above the Schwarzschild fitting line, enabling thus possibility of positive fitting for microquasars. However, in the range of allowed values of the wormhole parameters the fitting is not allowed.

\subsection{Active galactic nuclei fitting by the relativistic precession model}
In the case of the fitting of data from the active galactic nuclei we use the same procedure and the same values of dimensionless wormhole parameters as in the case of the microquasars. In this case the character of the fitting lines related to wormholes excludes the positive fitting in the case of active galactic nuclei, as the data are located under the region of fitting by the Kerr black holes, while the wormhole lines related to the relativistic precession variant of the geodesic model are located above the Schwarzschild line. 

\begin{figure}[ht] 
	\includegraphics[width=\linewidth]{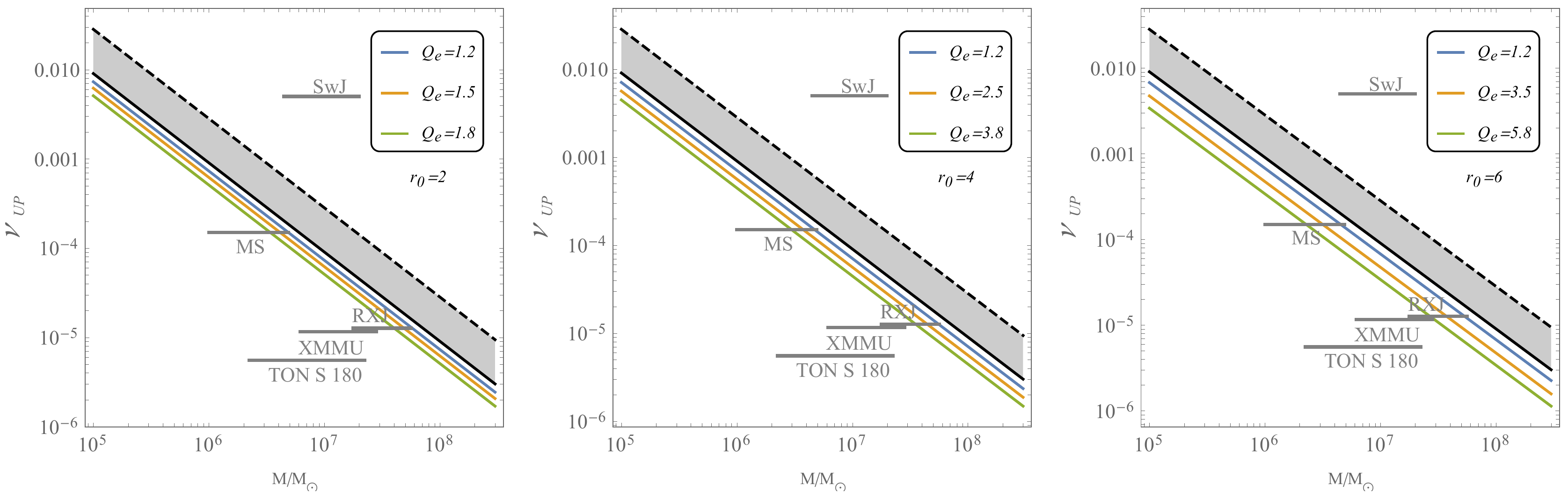}
	\caption{Fit of upper frequencies of 3:2 ratio to mass estimate of assumed supermassive black holes in active galactic nuclei for various values of the parameters $r_0$ and $Q_e$ by using the epicyclic orbital resonance model. The sources with unknown estimate of spin listed in Table \ref{t:tab1} (cf. \cite{Smi-Tan-Wag:2021:ApJ:}).}
	\label{f:f9}
\end{figure}
\begin{figure}[ht] 
	\includegraphics[width=\linewidth]{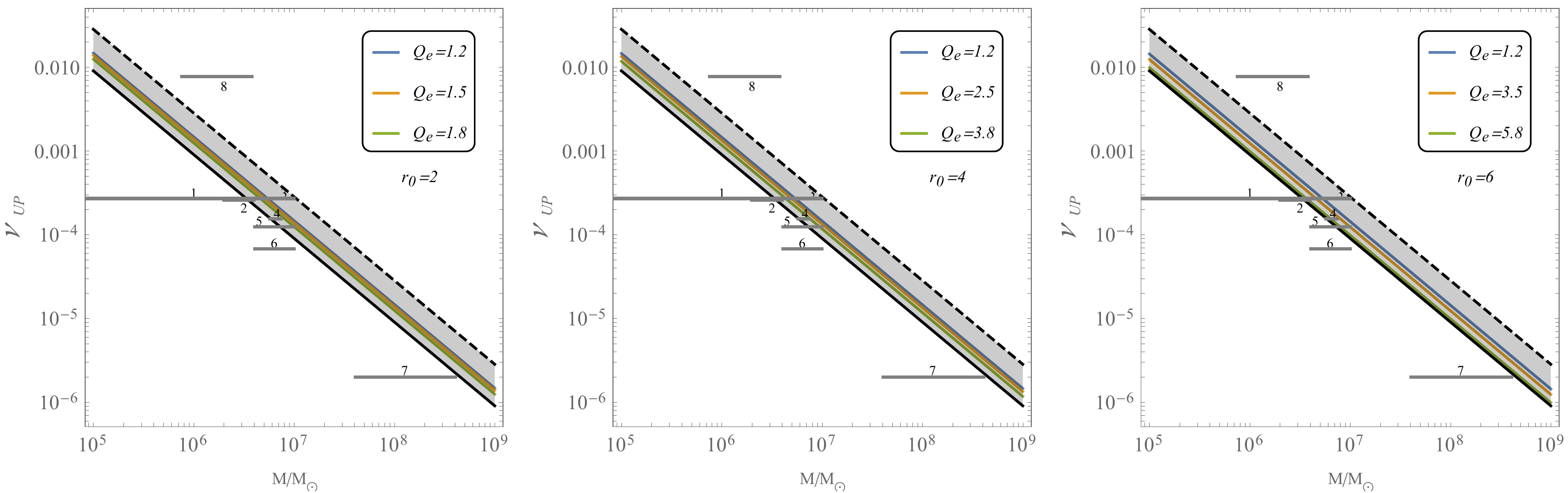}
	\caption{Fit of upper frequencies of 3:2 ratio to mass estimate of assumed supermassive black holes in active galactic nuclei for various values of the parameters $r_0$ and $Q_e$ by using the epicyclic relativistic precession model. The sources with known estimate of spin listed in Table \ref{t:tab1} (cf.  \cite{Smi-Tan-Wag:2021:ApJ:}).}
	\label{f:f10}
\end{figure}
\begin{figure}[ht] 
	\includegraphics[width=\linewidth]{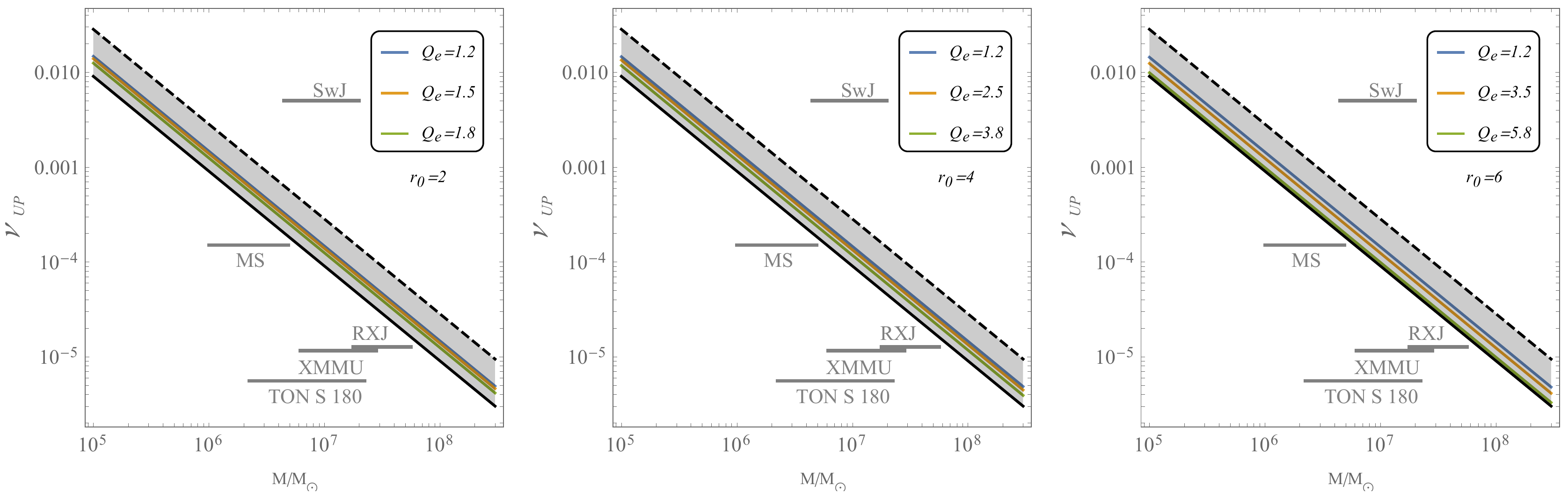}
	\caption{Fit of upper frequencies of 3:2 ratio to mass estimate of assumed supermassive black holes in active galactic nuclei for various values of the parameters $r_0$ and $Q_e$ by using the epicyclic relativistic precession model. The sources with unknown estimate of spin listed in Table \ref{t:tab1} (cf. \cite{Smi-Tan-Wag:2021:ApJ:}).}
	\label{f:f11}
\end{figure}

\begin{table*}[h]
\caption{Observations of QPOs around Supermassive Black Holes \cite{Smi-Tan-Wag:2021:ApJ:}}
\label{t:tab1}
\begin{tabular*}{\textwidth}{@{\extracolsep{\fill}}cllllllll@{}}
\hline
Number  &  Name & BH spin & log $M_\mathrm{BH}$ [$M_\odot$] & $f_\mathrm{UP}$ [Hz]  \\
\hline
1  &  MCG-06-30-15	&	$>0.917$	&	$6.20^{+0.09}_{-0.12}$	&	$2.73\times10^{-4}$	\\
2  &  1H0707-495	&	$>0.976$	&	$6.36^{+0.24}_{-0.06}$	&	$2.6\times10^{-4}$	\\ 
3  &  RE J1034+396	&	0.998	&	$6.0^{+1.0}_{-3.49}$	&	$2.7\times10^{-4}$	\\
4  &  Mrk 766	&	$>0.92$	&	$6.82^{+0.05}_{-0.06}$	&	$1.55\times10^{-4}$	\\
5  &  ESO 113-G010	&	0.998	&	$6.85^{+0.15}_{-0.24}$	&	$1.24\times10^{-4}$	\\
6  &  ESO 113-G010b	&	0.998	&	$6.85^{+0.15}_{-0.24}$	&	$6.79\times10^{-5}$	\\
7  &  1H0419-577	&	$>0.98$	&	$8.11^{+0.50}_{-0.50}$	&	$2.0\times10^{-6}$	\\
8  &  ASASSN-14li	&	$>0.7$	&	$6.23^{+0.35}_{-0.35}$	&	$7.7\times10^{-3}$	\\
-  &  TON S 180	&	$< 0.4$	&	$6.85^{+0.5}_{-0.5}$	&	$5.56\times10^{-6}$\\
-  &  RXJ 0437.4-4711	&	-	&	$7.77^{+0.5}_{-0.5}$	&	$1.27\times10^{-5}$	\\
-  &  XMMU J134736.6+173403	&	-	&	$6.99^{+0.46}_{-0.20}$	&	$1.16\times10^{-5}$	\\
-  &  MS 2254.9-3712	&	-	&	$6.6^{+0.39}_{-0.60}$	&	$1.5\times10^{-4}$ \\
-  &  Sw J164449.3+573451	&	-	&	$7.0^{+0.30}_{-0.35}$	&	$5.01\times10^{-3}$		\\
\hline
\end{tabular*}
\end{table*}
\begin{table*}[h]
\caption{Best fits of  parameters on observed Supermassive Black Holes with QPO signature}
\label{t:tab2}
\scalebox{1}{
\begin{tabular}{ccccc}
\hline
\multicolumn{1}{|c|}{\multirow{2}{*}{Name}} & \multicolumn{2}{c|}{Epicyclic model}                             & \multicolumn{2}{c|}{Relativistic prcession model}                             \\ \cline{2-5} 
\multicolumn{1}{|c|}{}                      & \multicolumn{1}{c|}{$r_0$} & \multicolumn{1}{c|}{$Q_e$} & \multicolumn{1}{c|}{$r_0$} & \multicolumn{1}{c|}{$Q_e$} \\ \hline
(1) MCG-06-30-15  & $ 2-4$ ($6$) & $1.8-3.8$, ($3.5-5.8$)               & -  & -  \\
(2) 1H0707-495   & $2-6$ & $1.2-3.5 $ & $6$  & $5.8$ \\
(3) RE J1034+396  & $2-6$ & $1.2-5.8$  & $2-6$ & $1.2-5.8$  \\
(4) Mrk 766  & -  & - & $4-6$ &  $3.8-5.8$ \\
(5) ESO 113-G010a & $2-4$  & $1.2-3.5$ & $2-6$ & $1.8-5.8$  \\
(6) ESO 113-G010b & $2-6$  & $1.5-5.8$ & - & - \\
(7) 1H0419-577   & $2-6$  & $1.2-5.8$  & - & - \\
(8) ASASSN-14li   & -   & - & - & - \\
TON S 180  & - & - & - & - \\
RXJ 0437.4-4711   & $2-6$ & $1.2-5.8$ & - & - \\
XMMU J134736.6+173403   & $6 $ & $5.8$  & - & - \\
MS 2254.9-3712   &  $2-6$ & $1.2-5.8$ & -  & -   \\
Sw J164449.3+573451     & -   & -    & -  & -  \\ \hline
\end{tabular}
}
\end{table*}
\FloatBarrier

\section{Conclusions}

We have studied the epicyclic oscillatory motion in the field of the EDM wormholes and give their frequencies, along with the frequency of the orbital (Keplerian) motion, as related to the distant static observers. We applied these frequencies in the framework of the geodesic model of HF QPOs, namely we used two variant, the epicyclic resonance, and the relativistic precession. These were applied to fitting the observational data from the three known microquasars, and to the group of active galactic nuclei discussed in \cite{Smi-Tan-Wag:2021:ApJ:}. 

We have demonstrated that for microquasars the observational data exclude existence of EDM wormholes with reasonable low values of the two free parameters. This is true for both the used variants of the geodesic model, the restriction is of absolute character for the epicyclic resonance variant. The relativistic precession variant could allow the fitting of data for enlarged region of the free parameters. 

On the other hand, in the fitting procedure of the data related to the active galactic nuclei, it was clearly demonstrated that the epicyclic resonance model enables positive results at least for many of the considered sources -- but not for all of them. By comparison, in Simpson-Visser meta geometry describing wormholes, most of these sources are also covered, but only at the expense of an enormous increase in (probably unrealistic) the length-scale parameter $ l $. Here, however, we cover most sources even without extreme parameter values. We summarize the results of the fitting procedure in Table \ref{t:tab2}. On the other hand, the relativistic precession model due to its properties does not allow for positive fitting the data from active galactic nuclei. 

Of course, if the rotational effects will be considered, one can expect some modification of the parameters resulting from the fitting, however, we are not expecting some fundamental negative shifts in the fitting of the sources. What can be expected is a more precise fitting and estimation of the parameters.

We can thus conclude that at least in some of the active galactic nuclei the existence of supermassive EDM wormholes cannot be excluded due the HF QPO data, while we can exclude them in microquasars, if reasonable values of the parameters  are considered. 

\bibliographystyle{abbrvnat}
\bibliography{references}

\end{document}